\documentclass[
aps,
prd,
showpacs,
preprint,
tightenlines,
amsmath
]{revtex4}
\newcommand{\ud}{\mathrm{d}}
\newcommand{\eq}[1]{Eq.(\ref{#1})}

\newcommand{\sqw}{\sin^2{\theta_W}}
\newcommand{\gevsq}{\,\mathrm{GeV}^2}

\begin{document}

\title{Paschos--Wolfenstein Relationship for Nuclei
and the NuTeV $\sqw$ Measurement}

\author{S. A.~Kulagin}
\email[]{kulagin@ms2.inr.ac.ru}
\affiliation{Institute for Nuclear Research, 117312
Moscow, Russia}


\begin{abstract}
We discuss nuclear effects in the Paschos--Wolfenstein relationship in the
context of extraction of the weak mixing angle. We point out that the
neutron excess correction to the Paschos--Wolfenstein relationship for a
neutron-rich target is negative and large on the scale of experimental
errors of a recent NuTeV measurement. We found a larger neutron excess
correction to the Paschos--Wolfenstein relationship for total cross sections
than that discussed by the NuTeV collaboration. Phenomenological
applications of this observation are discussed in the context of the NuTeV
deviation. Uncertainties in the neutron excess correction are estimated.
Effects due to Fermi motion, nuclear binding, and nuclear shadowing are also
discussed in the context of total cross sections.
\end{abstract}

\pacs{13.15.+g, 13.60 Hb, 24.85.+p}
\maketitle

The NuTeV collaboration recently reported the results of the measurement of
the weak mixing angle in deep inelastic neutrino and antineutrino
scattering from a heavy target \cite{nutev-prl}. The NuTeV value of
$\sqw=0.2277\pm 0.0013(\text{stat.})\pm 0.0009(\text{syst.})$
turned out to be significantly larger than that derived from a global
standard model fit to other electroweak measurements, $0.2227\pm 0.0004$
\cite{lep}. The discussion of possible uncertainties and physics behind
this discrepancy can be found in ref.\cite{DFGRS02}. However, before
speculating on possible new physics, one should first worry about
``standard'' effects and uncertainties.

A useful tool, employed by
the  NuTeV collaboration to derive $\sqw$, is the
Paschos--Wolfenstein relationship (PW) \cite{PW73}
\begin{eqnarray}
\label{pw}
R^- &=&
\frac{
\sigma^\nu_{\text{NC}}-\sigma^{\bar\nu}_{\text{NC}}
}
{
\sigma^\nu_{\text{CC}}-\sigma^{\bar\nu}_{\text{CC}}
}
=\frac12 -\sqw,
\end{eqnarray}
where $\sigma^\nu_{\text{NC}}$ and $\sigma^\nu_{\text{CC}}$ are the deep
inelastic neutrino cross sections for neutral current (NC) and
charged-current (CC) interactions, and $\sigma^{\bar\nu}_{\text{NC}}$ and
$\sigma^{\bar\nu}_{\text{CC}}$ are corresponding antineutrino cross sections
\footnote{NuTeV experiment was primarily designed to measure $\sqw$
using the Paschos--Wolfenstein relationship \cite{nutev-nuint01}. However,
it must be noted that the NuTeV collaboration does not directly measure the
cross sections and the ratio $R^-$, but rather measures ``ratios of
experimental candidates within kinematic criteria and compares this to a
Monte Carlo simulation'' \cite{nutev-prd}.}.
%

In the world without heavy
quarks and with exact isospin symmetry, \eq{pw} is an exact relation for the
isospin zero target for both the total and differential cross sections. The
validity of this relationship is solely based on the isospin symmetry
\cite{PW73}. Therefore, \eq{pw} also holds for a nuclear target,
provided that the nucleus is in the isoscalar state. In particular, this
means that various strong interaction effects, including nuclear effects,
must cancel out in the ratio (\ref{pw}). However, in the real world \eq{pw}
must be corrected for the $s$- and $c$-quark effects. Furthermore, the
targets used in neutrino experiments are usually non-isoscalar nuclei, and
\eq{pw} must also be corrected for non-isoscalarity effects, as well as for
other nuclear effects such as Fermi motion, binding, and nuclear shadowing.

Nuclear effects in the context of $\sqw$ were recently discussed in
refs.\cite{mt02,KSY02,Kumano}. It was suggested in
ref.\cite{mt02} that a difference between the nuclear shadowing in NC and CC
interactions may account for the discrepancy in NuTeV's measurement of
$\sqw$, although no specific calculation of this effect was given. The impact
of nuclear effects on the extraction of $\sqw$ from data
was discussed in ref.\cite{KSY02} in terms of a phenomenological
parameterization of nuclear effects in parton distributions derived from
charged lepton deep inelastic scattering.
Phenomenological studies of the
difference between nuclear effects in the $u$- and $d$-quark distributions
in the context of the PW relationship were presented in ref.\cite{Kumano}.

In this paper we address nuclear effects in the PW ratio for total cross
sections. We start from the discussion of the non-isoscalarity correction to
the PW ratio, since there appears to be some confusion about the sign and
the magnitude of this effect in the literature. Then we discuss Fermi
motion, nuclear binding, and nuclear shadowing corrections to the PW ratio.

\section{The PW relationship for generic target}

We discuss (anti)neutrino deep inelastic scattering in the
leading twist QCD approximation assuming that the four-momentum
transfer $Q$ is large enough. In this approximation the NC and CC structure
functions are given by well-known expressions in terms of quark and
antiquark distribution functions \cite{Hagiwara:fs}. In order to simplify
discussion of isospin effects, we consider the isoscalar,
$q_0(x)=u(x)+d(x)$, and isovector, $q_1(x)=u(x)-d(x)$, quark distributions
(for simplicity, we suppress the explicit notation for the $Q^2$ dependence
of parton distributions). The calculation of the NC and CC cross sections,
and the PW ratio in the QCD parton model is straightforward. A QCD radiative
correction to the PW ratio for the total cross sections was calculated in
ref.\cite{DFGRS02}. The result can be written as
%
\begin{eqnarray}
\label{pw:cor}
R^- =
\frac12 - s_W^2 
{}+\left[1-\frac73 s_W^2 +
        \frac{4\alpha_s}{9\pi}\left(\frac12-s_W^2\right)\right]
\left(
\frac{x_1^-}{x_0^-}-\frac{x_s^-}{x_0^-}+\frac{x_c^-}{x_0^-}
\right),
\end{eqnarray}
%
where $s_W^2=\sqw$
\footnote{
\eq{pw:cor} was derived in the tree approximation. Electroweak corrections
depend on the definition of the weak mixing angle. To be specific, the weak
mixing angle is defined here in the on-shell scheme, $s_W^2=1-M_W^2/M_Z^2$.},
$\alpha_s$ is the strong coupling, and $x_a^- =\int \ud x\,x(q_a-\bar q_a)$,
with $q_a$ and $\bar q_a$ the distribution functions of quarks and
antiquarks of type $a$
\footnote{The total cross sections involve the
integration of the structure functions over the full phase space of $x$
and $Q^2$. Therefore $\alpha_s$ and the moments $x_i^-$ of the parton
distributions are taken at some average scale $Q^2$, which has to be chosen
according to specific experimental conditions.}.
The subscripts 0 and 1
refer to the isoscalar $q_0$ and isovector $q_1$ quark distributions,
respectively. In the derivation of \eq{pw:cor} we expanded in
$x_{1}^-/x_0^-$ and $x_{s,c}^-/x_0^-$ and retained only linear corrections.
\eq{pw:cor} applies to any (not necessarily isoscalar) nuclear target. We
observe that the PW relationship (\ref{pw}) is corrected by the $C$-odd parts
of the isovector component in the target ($x_1^-$) and the strange and charm
components of the target's sea.

The isovector quark distribution $q_1$ vanishes in an isoscalar target,
provided that the isospin symmetry is exact. However, a correction due to the
quark--antiquark asymmetry in the nucleon (or nuclear) strange sea
is possible. In particular, a positive $x_s^-$
would move $s_W^2$ towards a standard model value \cite{DFGRS02}. However,
available phenomenological estimates are controversial even in the
sign of the effect: a shift in $s_W^2$, estimated in ref.\cite{DFGRS02}, is
$-0.0026$, however the NuTeV collaboration gives a positive shift of
$0.0020\pm 0.0009$ \cite{nutev-prd}.

\section{Non-isoscalarity correction in a heavy target}

Heavy nuclei, such as iron, have unequal number of neutrons ($N$) and
protons ($Z$). The ground state of such nuclei is not isospin zero state
but rather a mixture of different isospins. Therefore, the isovector quark
distribution is finite in such nuclei that causes a finite non-isoscalarity
correction to structure functions and cross sections. This correction
turns out to be large in the context of the NuTeV experiment, and,
therefore, must carefully be taken into account.

In order to understand this
effect, we first neglect other nuclear effects and view the neutrino
scattering off a nucleus as incoherent scattering off bound protons and
neutrons at rest. In this approximation, the nuclear distribution of
partons of type $a$ is the sum of those for bound protons and neutrons
\begin{eqnarray}\label{nuke:NZ}
q_{a{/}\mathrm{A}} &=& Z q_{a{/}\mathrm{p}}+ N q_{a{/}\mathrm{n}}
= \frac{A}{2}\left(q_{a/\mathrm{p}}+q_{a/\mathrm{n}}\right)
+ \frac{Z-N}{2}\left(q_{a/\mathrm{p}}-q_{a/\mathrm{n}}\right),
\end{eqnarray}
where $A=Z+N$.
We now apply this relation to the isoscalar and the isovector quark
distributions. Assuming exact isospin symmetry, we have
$q_{0/\mathrm{p}}=q_{0/\mathrm{n}}$ and
$q_{1/\mathrm{p}}=-q_{1/\mathrm{n}}$. For nuclear parton distributions per
one nucleon we then obtain
\begin{subequations}
\label{nuke:01}
\begin{eqnarray}
\label{nuke:0}
A^{-1}q_{0/\mathrm{A}} &=& q_{0/\mathrm{p}},\\
A^{-1}q_{1/\mathrm{A}} &=& -\delta N q_{1/\mathrm{p}},
\label{nuke:1}
\end{eqnarray}
\end{subequations}
where we introduced a fractional excess of neutrons $\delta N=(N-Z)/A$.
Similar equations can readily be written for antiquark distributions.

Eqs.(\ref{nuke:01}) suggest a \emph{negative} neutron excess correction to
$R^-$ in a neutron-rich target. Indeed, using Eqs.(\ref{nuke:01}), we find
from \eq{pw:cor} that the correction is
\begin{equation}
\label{dn-cor}
\delta R^-=-\delta N\,\frac{x_1^-}{x_0^-}
        \left[1-\frac73 s_W^2 +
        \frac{4\alpha_s}{9\pi}\left(\frac12-s_W^2\right)\right],
\end{equation}
where $x_1^-$ and $x_0^-$ are taken for the proton.

The NuTeV collaboration takes into account the non-isoscalarity correction
in the analysis and discusses this effect on $R^-$
\cite{nutev-nuint01,nutev-prd}.
However, \eq{dn-cor} is different from the corresponding NuTeV equation
(Eq.(7) in ref.\cite{nutev-prd} and Eq.(9) in ref.\cite{nutev-nuint01}).
First, it must be noted that the NuTeV equation
has the wrong sign of the $\delta N$
term. Nevertheless, the NuTeV non-isoscalarity
correction is eventually negative \cite{nutev-pc}. Therefore, in the
following discussion we assume a correct sign in Eq.(7) in
ref.\cite{nutev-prd}.
Second, \eq{pw:cor} involves only $C$-odd terms, hence the factor
$x_1^-/x_0^-$. The corresponding factor in
refs.\cite{nutev-nuint01,nutev-prd} is $(x_u-x_d)/(x_u+x_d)$, where
$x_a=\int \ud x\, xq_a(x)$ for the quark distribution $q_a$ in the proton.
This factor has a mixed $C$-parity and for this reason is not allowed in
\eq{dn-cor}. Furthermore, \eq{dn-cor} includes the $\alpha_s$ correction,
which was not taken into account in refs.\cite{nutev-nuint01,nutev-prd}.

In order to understand the magnitude of the non-isoscalarity correction to
$R^-$ and compare it to that in refs.\cite{nutev-nuint01,nutev-prd}, we first
neglect the
$\alpha_s$ correction and compute $x_1^-/x_0^-= 0.43$ and
$(x_u-x_d)/(x_u+x_d)= 0.34$ using CTEQ5 parton distributions in
$\overline{\text{MS}}$ scheme \cite{cteq5} evaluated at $Q^2=20\gevsq$, an
average $Q^2$ in the NuTeV experiment. For the neutron excess we use the
value $\delta N=0.0567$ reported by NuTeV \cite{nutev-prd}. Using
$s_W^2\simeq0.22$ and collecting all factors in \eq{dn-cor}, we obtain the
neutron excess correction to $R^-$ about $-0.012$. The corresponding
correction computed using Eq.(7) in
ref.\cite{nutev-prd} is  $-0.0094$. This correction is smaller in
magnitude because the factor $(x_u-x_d)/(x_u+x_d)$ is smaller than the factor
$x_1^-/x_0^-$ by about 25\%. If we simply apply the difference, $-0.0026$,
to the NuTeV central value of $\sqw$, we get a new value, $0.2251$, which is
 now about $1.5\sigma$ away from the standard model value.
The NuTeV collaboration also reported the neutron excess correction to
be $-0.0080$ \cite{nutev-nuint01,nutev-pc}. In this case the additional
shift in the NuTeV $\sqw$ is $-0.004$, and
the discrepancy between the NuTeV and the standard model value of $\sqw$
practically disappears.
However, it must be noted that although the non-isoscalarity correction
of $-0.008$ was discussed in ref.\cite{nutev-nuint01} in the context of
$R^-$, it does not directly apply to the ratio of cross sections. The NuTeV
collaboration explains that this is the shift in the NuTeV $\sqw$ that takes
into account experimental background and the cuts \cite{nutev-pc}.

The $\alpha_s$ correction to $R^-$ is also negative, but small compared to
the leading term. Using the NLO $\alpha_s=0.2161$ at $Q^2=20\gevsq$, we find
the $\alpha_s$ correction to $R^-$ about $-0.0002$.

Let us now discuss possible uncertainties in the non-isoscalarity
correction.
According to NuTeV \cite{nutev-prl}, the uncertainty in this
correction to $\sqw$ is $0.00005$. Apparently, this is due to the
uncertainty in the neutron excess in the target
\footnote{Using the $0.05\%$ uncertainty in $\delta N$ \cite{nutev-prd}
and \eq{dn-cor}, we obtain the uncertainty in $\sqw$ about $0.0001$.}.
However, this is not the only source of uncertainties in the $\delta N$
correction. Indeed, the ratio $x_1^-/x_0^-$ is subject to theoretical and
experimental uncertainties. This ratio depends on the set of parton
distributions as well as on the order of perturbation theory to which the
analysis is performed
\footnote{The NuTeV collaboration
uses its own parton distributions obtained in the leading order fit to
their CC data. We comment in this respect that, since the NuTeV
collaboration aims at precize determination of the weak mixing angle, the
inclusion of NLO corrections to parton distributions appears as a necessary
improvement of the analysis.}. For example, the ratio $x_1^-/x_0^-$
calculated with the MRST parton distributions \cite{MRST} at the same value
of $Q^2$ is very similar to that of CTEQ5. However, $x_1^-/x_0^-=0.445$ for
Alekhin's parton distributions \cite{a02}.
Furthermore, the parton distributions are known with some error. This error
was estimated in a recent analysis in ref.\cite{a02}. Using the parton
distributions of ref.\cite{a02}, we have the estimation on the uncertainty in
the ratio $\delta(x_1^-/x_0^-)\simeq0.04$. This results in the uncertainty in
$R^-$ about $0.001$, which is the order of magnitude larger than the
uncertainty in the neutron excess in the target.
As an illustration, we treat this uncertainty as an independent theoretical
uncertainty in $\sqw$ and add it in quadrature to the NuTeV error. Then we
find some $1.2\sigma$ distance between the standard model value and the
corrected value of NuTeV $\sqw$ discussed above.
We also comment that the variations of the
parton distributions with $Q^2$ introduce
additional uncertainty, which is hard to access in the present analysis.

\section{Fermi motion and nuclear binding corrections}

We briefly discuss other nuclear effects on $R^-$. In order to sort out
different effects, in the present discussion we do not consider explicit
violations of the isospin symmetry. We recall that the isospin symmetry
requires nuclear effects to cancel out in the $R^-$ ratio for the isoscalar
target. For a non-isoscalar nucleus, \eq{dn-cor} was derived neglecting the
effects of Fermi motion and nuclear binding. However, it is rather obvious
that these effects and the non-isoscalarity correction must be considered in
a unified approach. In such an approach Eq.(\ref{nuke:NZ}) must be replaced
by a convolution of quark distributions with nuclear distributions of bound
protons and neutrons \cite{binding}. Note that in the approximation in which
the proton and neutron distributions are identical, the Fermi motion and
nuclear binding effects cancel out in the ratio $x_1^-/x_0^-$ \footnote{
However, it must be emphasized that this cancellation applies to the total
cross sections. For the ratio of differential cross sections (or for the
quark distributions) this effect remains finite even in the approximation in
which the distribution functions of bound protons and neutrons are
identical. }.
Therefore, a possible correction comes through the difference between the
proton and neutron distribution functions and is likely to be small. One
source of the correction is a nonlinear dependence of the nucleon
distribution functions on the number of bound particles. In heavy nuclei
with $Z\not=N$ this effect results in different nucleon distribution
functions in the isovector and isoscalar channels even if no violation of
the isospin symmetry is admitted \footnote{
It is interesting to note that this effect may mask the effects of isospin
violation in quark distribution functions.
It must be also commented that possible difference between nuclear
modifications of $u$- and $d$-quark distributions was discussed in
ref.\cite{Kumano} in the context of the NuTeV deviation in terms of a
phenomenological approach. }. Because of this effect the ratio $x_1^-/x_0^-$
receives a correction factor of $1-\frac29T/M$, where $T$ is the average
kinetic energy of the bound nucleon and $M$ the nucleon mass \cite{kulagin}.
For the iron nucleus this factor is $0.993$ that gives only a small
correction to $R^-$.

\section{Nuclear shadowing effect}

We now discuss nuclear shadowing effect in the context of the
isoscalar and isovector quark distributions.
The nuclear shadowing corrections to the quark distributions can be written
as
\begin{subequations}
\label{sh:01}
\begin{eqnarray}
\label{sh:0}
A^{-1}q_{0/\mathrm{A}} &=& q_{0/\mathrm{p}}+\delta_{\text{sh}}q_0,\\
A^{-1}q_{1/\mathrm{A}} &=& -\delta N q_{1/\mathrm{p}}+\delta_{\text{sh}}q_1.
\label{sh:1}
\end{eqnarray}
\end{subequations}
Similar equations can be written for antiquark distributions.
The nuclear
shadowing effect in the isoscalar $C$-even ($q_0^+=q_0+\bar q_0$) and
$C$-odd ($q_0^-=q_0-\bar q_0$) quark distributions was studied in
ref.\cite{ku98}, where a different magnitude of the shadowing effect in
$q_0^+$ and in $q_0^-$ was observed. In particular, it was argued that the
relative shadowing effect in the $C$-odd quark distribution is enhanced
compared to that in the $C$-even distribution (see also \cite{nuF}).
In order to estimate the shadowing effect in the total cross sections, we
apply the approach of ref.\cite{ku98} and calculate the shadowing corrections
to the average quark light-cone momentum, $\delta_{\text{sh}}x_0$, in the
$C$-even  and $C$-odd  distributions. For the iron
nucleus the result is $\delta_{\text{sh}}x_0^+/x_0^+=-0.01$ and
$\delta_{\text{sh}}x_0^-/x_0^-=-0.002$. We observe that the shadowing
effect in the total cross sections is significantly bigger in the $C$-even
channel, in spite of the enhancement of the relative nuclear shadowing
effect in the $C$-odd quark distributions.
This, paradoxical from the first glance, result can be
understood by observing that the $C$-odd cross section is saturated by the
valence region at large $x$, where the nuclear shadowing effect is
small. On the other hand, a large part of the $C$-even cross section comes
from the sea region, where the nuclear shadowing effect is essential.

Since the isovector distribution must vanish in the isoscalar
nucleus, the shadowing effect in the isovector channel appears at least in
the order $\delta N$. This effect has not been quantitatively studied yet.
In our estimates we neglect the shadowing correction to $x_1^-$, leaving
this interesting question to further studies. Then,
the nuclear shadowing correction reduces to the renormalization of
the ratio $x_1^-/x_0^-$ by the factor of $1.002$ for the iron nucleus.

Combining the nuclear shadowing effect with the Fermi motion and nuclear
binding corrections to $R^-$, we observe a partial cancellation between
these effects. For the iron nucleus the resulting correction factor to the
ratio $x_1^-/x_0^-$ is $0.995$ that gives a negligible correction to $R^-$.

\section{Summary}

We have studied nuclear effects in the PW relationship for total
cross sections. We observed that for a neutron-rich target (such as iron,
used in the NuTeV experiment) the neutron excess correction to the PW
relationship is negative and large on the scale of experimental errors. We
found a larger neutron excess correction than that discussed by the NuTeV
collaboration in the context of $R^-$. The ratio $R^-$ is rather sensetive
to the value of the neutron excess correction and a consistent treatment of
this correction may well explain at least a large part of the NuTeV
deviation.

The uncertainties in nuclear non-isoscalarity correction were discussed. We
found that the uncertainty is dominated by the uncertainty in the parton
distributions leading to the variation in $R^-$ about $0.001$ that should be
considered as an additional source of theoretical uncertainty in $\sqw$.

We discussed the effects of Fermi motion, nuclear binding and nuclear
shadowing on the ratio $R^-$ for the total cross sections. We point out that
nuclear effects on $R^-$ vanish for the isoscalar nucleus and appear in the
first or higher order in $\delta N$. For this reason these effects are small.
Furthermore, we observed a partial cancellation between Fermi motion and
nuclear shadowing effects in the ratio $R^-$.

However, this cancellation of nuclear effects in the ratio $R^-$ for the
total cross sections does not, of course, mean that these effects
identically cancel out in the ratio of differential cross sections. Since
the NuTeV experiment does not measure the total cross sections, some nuclear
effects which cancel out in $R^-$ remain in actual experimental observables.
Therefore, in order to clarify the impact of nuclear effects on the NuTeV
result, it is necessary to explicitly take them into account in the
analysis.

\begin{acknowledgments}

This work was supported in part by the RFBR project no. 00-02-17432.
I am grateful to S.I. Alekhin, A.L. Kataev, S. Liuti, W.
Melnitchouk, R. Petti, V.A. Rubakov, and G.P. Zeller for useful discussions
and to the Theory Division of CERN, where a part of this work was done, for
a warm hospitality.

\end{acknowledgments}

\end{document}